\documentclass[a4paper,11pt]{article}
\pdfoutput=1 

\usepackage{jheppub} 

\usepackage[T1]{fontenc} 
\usepackage{comment}
\usepackage{appendix}
\usepackage[dvipsnames]{xcolor}

\makeatletter
\gdef\@fpheader{\normalfont https://doi.org/10.1007/JHEP06(2022)159}
\makeatother

\title{Mixing Particle Production for Relaxion Mechanism}

\author[a,b]{Tanech Klangburam}
\author[a]{Areef Waeming}
\author[a,c]{Predee Tantirangsri}
\author[a,b]{Daris Samart}
\author[a,b]{Chakrit Pongkitivanichkul}

\affiliation[a]{Khon Kaen Particle Physics and Cosmology Theory Group (KKPaCT),\\ Department of Physics, Faculty of Science, Khon Kaen University, 123 Mitraphap Rd.,\\ Khon Kaen, 40002, Thailand}
\affiliation[b]{National Astronomical Research Institute of Thailand, Chiang Mai 50180, Thailand}
\affiliation[c]{Faculty of Science, Mahidol University, 272 Rama VI Rd., Ratchathewi,\\ Bangkok 10400, Thailand}

\emailAdd{klangburam.t@gmail.com}
\emailAdd{reef.waeming@gmail.com}
\emailAdd{predeetan@gmail.com}
\emailAdd{darisa@kku.ac.th}
\emailAdd{chakpo@kku.ac.th}

\abstract{We consider the production of two heavy gauge bosons as a relaxation stopping mechanism. In this work, we analyse the conditions for a tachyonic mode for a linear combination of gauge bosons and show that the criteria are significantly different than the single gauge boson case. Moreover, the implementation of the mechanism on the $U(1)'$ model is demonstrated. We discuss various constraints for the relaxion mechanism. The phenomenology of the heavy gauge boson is also explored. We finally show a benchmark point of parameter space considering all constraints from relaxion and the $U(1)'$ mixing sector.}

\keywords{Axions and ALPs, Cosmology of Theories BSM, Hierarchy Problem, New Light Particles}

\begin{document} 
\maketitle
\flushbottom

\section{Introduction}

The electroweak hierarchy problem has been one of the main motivations for physics beyond the standard model. The hierarchy problem is linked to a theoretical aspect of the quantum field theory framework in which the measured Higgs mass requires a very delicate balance between two different scales, namely the electroweak scale ($\sim 10^3$ GeV) and the Planck scale ($\sim 10^{19}$ GeV). In order to resolve this unnaturalness, one often resorts to either introducing a powerful symmetry such as supersymmetry or falling back to an anthropic argument. However, with an increasing effort to find evidence of new particles, models with supersymmetry are currently forced to reside in the fine-tuned corner of the parameter space.

Recently, there has been an alternative class of solutions to the hierarchy problem called axion relaxation, aka relaxion. The original model \cite{Graham:2015cka} utilises the dynamical evolution of the Higgs mass in an early universe in order to scan for the correct value of the electroweak scale. The idea can be summarised as follows. Consider the effective Lagrangian of the relaxion, $\phi$ and Higgs sector
\begin{equation}
    \mathcal{L}\supset (\Lambda^2 - \kappa \phi) h^{\dagger}h + \kappa \phi \Lambda^2 + \ldots,
\end{equation}
where $\Lambda$ is the UV cutoff energy and $\kappa$ is the dimensionful parameter coming from a spurion field which break shift symmetry of the axion $\phi$. Due to the linear slope of the potential, the relaxion rolls down during an early stage of the universe. Then this causes an effective Higgs mass, $-\Lambda^2 + \kappa \phi$, to change until the sign is switched and the Electroweak symmetry breaking is turned on. The Higgs vacuum expectation value (vev) then creates the periodic potential for the relaxion
\begin{equation}
V(\phi) \supset \Lambda_c^4 \cos\left(\frac{\phi}{f}\right),
\end{equation}
where its amplitude, $\Lambda_c$, grows with the Higgs vev. The increasingly higher barrier then dissipates the kinetic energy and traps relaxion in one of the minima. The result is that an effective Higgs mass is naturally fixed at a small value comparing to the cutoff scale, therefore the hierarchy problem is solved by the dynamics of the relaxion.

However, the original model faces a few theoretically unsatisfactory aspects. Since relaxion has to travel a large field distance, the simplest implementation of the model generically produces a large amount of inflation and a low inflation scale. 
In order to decouple relaxion from the dynamics of inflation, particle production has been proposed as an alternative to the relaxion stopping mechanism \cite{Hook:2016mqo}. The anomaly term for a $U(1)$ gauge boson is allowed by shift symmetry
\begin{equation}
    \mathcal{L} \supset -\frac{\phi}{4f} F_{\mu\nu} \widetilde{F}^{\mu\nu},
\end{equation}
where $F_{\mu\nu}$ is a field strength tensor. The anomaly coupling leads to an exponential production of gauge bosons if theirs mass is light enough to be produced, i.e., the tachyonic mode of gauge bosons is allowed when
\begin{equation}
    \dot{\phi} > f m_A,\label{eq:ppcond1}
\end{equation}
where $m_A$ is the gauge boson mass. To employ this mechanism in an early universe, an effective Higgs mass, $-\Lambda^2 + \kappa \phi$, is assumed to start off with a large and negative value. Then as the relaxion rolls down the potential, the vev of Higgs decreases until the condition Eq.(\ref{eq:ppcond1}) is met and the gauge boson is exponentially produced. The mechanism has been shown to efficiently dissipate relaxion's kinetic energy and trap it in one of the minima \cite{Anber:2009ua,Kofman:2004yc}. In this paper, we will focus on particle production as the main stopping mechanism.

This solution to the electroweak hierarchy problem sparks a huge interest in various fields in the literature. Various models using relaxion mechanism are proposed in \cite{Espinosa:2015eda,Antipin:2015jia,Hardy:2015laa,Batell:2015fma,Matsedonskyi:2015xta,Evans:2016htp,Huang:2016dhp,Agugliaro:2016clv,Lalak:2016mbv,You:2017kah,Evans:2017bjs,Batell:2017kho,Ferreira:2017lnd,Matsedonskyi:2017rkq,Davidi:2017gir,Fonseca:2017crh,Son:2018avk,Fonseca:2018xzp,Davidi:2018sii,Fonseca:2018kqf,Abel:2018fqg,Wang:2018ddr,Gupta:2019ueh,Fonseca:2019aux,Ibe:2019udh,Kadota:2019wyz,Fonseca:2019ypl,Fonseca:2019lmc,Banerjee:2020kww,Domcke:2021yuz} 
where the model building aspects can be found in \cite{Gupta:2015uea,Abel:2015rkm,Choi:2015aem,Ibanez:2015fcv,Hebecker:2015zss,McAllister:2016vzi,Fonseca:2016eoo,Nelson:2017cfv,Gupta:2018wif}. The studies on inflation and reheating in an early universe have been done in \cite{Patil:2015oxa,Jaeckel:2015txa,Marzola:2015dia,DiChiara:2015euo,Tangarife:2017rgl,Choi:2016kke}. The signatures in experiments and cosmology are studied in \cite{Kobayashi:2016bue,Choi:2016luu,Flacke:2016szy,Beauchesne:2017ukw,Frugiuele:2018coc,Banerjee:2018xmn,Banerjee:2019epw,Barducci:2020axp,Banerjee:2021oeu}.

As the standard model is extended, it is more typical for an axion to receive more than one non-perturbative effect. For example, in the String/M theory context, the number of string/membrane instantons present in any compactification is larger than the number of axions as they are required for stabilisation of the axion/moduli potential. We are interested in the case that the dark/sequestered sector contains a gauge symmetry, which is the source of the extra contribution to the axion potential.

Motivated by a gauge symmetry in the dark sector, we consider the model of relaxion coupled with $U(1)'$ from dark sector \cite{Holdom:1985ag,DelAguila:1995fa,Langacker:2008yv,Pospelov:2008zw}. We will show that the $U(1)'$ mixing with the standard model gauge bosons leads to a new threshold for particle production mechanism. The result lifts the cut off scale by many orders of magnitude compared to typical models. The extra non-perturbative effect will be added to the model. The model will also give rise to an interesting phenomenology of the $U(1)'$ gauge boson.

Besides the particle production, another stopping mechanism called "Axion fragmentation" has been discussed \cite{Fonseca:2019lmc,Fonseca:2019ypl}. The effect has been overlooked in the original relaxion models as it is always present, even in the simplest realisation. As an axion rolling down the potential and passing through many wiggles, the axion fluctuation can be described by the Mathieu equation. The parametric resonance effect for some of the momentum modes could produce an exponentially large number of particles, providing an extra source of friction to the relaxion stopping mechanism. Since the exponential growth of the perturbations occurs along the distance that an axion travels, the axion fragmentation effect dominates over the gauge boson production, and the vacuum expectation value will be much closer to the cut-off scale. Therefore, the axion fragmentation in the $U(1)'$ coupled model of relaxion will be investigated.

The paper is organised as follow. The particle production from Abelian gauge bosons will be discussed in section \ref{sec:pp}. The implementation of the model on the axion relaxation is shown in section \ref{sec:model} where the axion fragmentation effect will also be analysed. In section \ref{sec:result}, examples of the parameters regarding all constraints are shown. Our conclusions are drawn in the last section \ref{sec:conclusion}.


\section{Particle Production from Mixing $Z$ with $Z'$ \label{sec:pp}}
In this section, we construct a model and study the particle production mechanism from the relaxion coupling with $Z$ and $Z'$. The Lagrangian including the relaxion anomalous coupling can be written as
\begin{align}
 \mathcal{L}=&-\frac{1}{4}Z_{a\mu\nu}Z_a^{\mu\nu} -\frac{1}{4}Z_{b\mu\nu}Z_b^{\mu\nu} -\frac{\epsilon}{2}Z_{a\mu\nu}Z_b^{\mu\nu} 
 -\frac{1}{2}M_a^2 Z_{\mu}Z^\mu -\frac{1}{2}M_b^2 Z'_{\mu}Z'^\mu \nonumber \\
 &-\frac{\phi}{4 f_a}Z_{a\mu\nu}\tilde{Z}_a^{\mu\nu} -\frac{\phi}{4 f_b}Z_{b\mu\nu}\tilde{Z}_b^{\mu\nu}, \label{eq:LZZ}
\end{align}
where the field $\phi$ is relaxion, and the $\epsilon$ is mixing parameter. $Z_i^{\mu \nu}$ are the field strength tensors where $i=a,b$ indicates the species of $U(1)_Y$ and $U(1)'$ gauge bosons respectively. The tilde represents the dual field, i.e., $\tilde{Z}_i^{\mu\nu}=\frac{1}{2}\epsilon^{\mu\nu\alpha\beta}Z_{i\alpha\beta}$. The kinetic terms can be diagonalised by
\begin{align}
    \begin{pmatrix}
    Z^\mu_a \\ 
    Z^\mu_b
    \end{pmatrix} =
    \begin{pmatrix}
    \frac{1}{\sqrt{1-\epsilon^2}} & 0 \\
    -\frac{\epsilon}{\sqrt{1-\epsilon^2}} & 1
    \end{pmatrix}\begin{pmatrix}
    Z^\mu\\
    Z'^\mu
    \end{pmatrix}.\label{eq:LaZ}
\end{align}
The canonical Lagrangian becomes
\begin{align}
    \mathcal{L}=&-\frac{1}{4}Z_{\mu\nu}Z^{\mu\nu} -\frac{1}{4}Z'_{\mu\nu}Z'^{\mu\nu}  -\frac{1}{2}M_a^2 Z_{\mu}Z^\mu -\frac{1}{2}M_b^2 Z'_{\mu}Z'^\mu \nonumber \\
    &-\frac{\left(\epsilon^2 f_a + f_b \right)\phi}{4 \left(1-\epsilon^2\right) f_a f_b}Z_{\mu\nu}\tilde{Z}^{\mu\nu} -\frac{\phi}{4 f_b}Z'_{\mu\nu}\tilde{Z'}^{\mu\nu}
    -\frac{\epsilon \phi}{2 \sqrt{1-\epsilon^2} f_b}Z_{\mu\nu}\tilde{Z}'^{\mu\nu},\label{eq:diaZ}
\end{align}
where $Z_{\mu\nu}$ and $Z'_{\mu\nu}$ are field strength tensors of $Z$ boson and  $Z'$ boson respectively. The equations of motion can be written in terms of two circular polarisations $Z_\pm$ and $Z'_\pm$ as
\begin{align}
    \ddot{Z}_\pm + \left( k^2 + M_a^2 \pm \frac{\left(\epsilon^2 f_a + f_b \right)k\dot{\phi}}{ \left(1-\epsilon^2\right) f_a f_b} \right)Z_\pm \pm \frac{\epsilon \dot{\phi}}{ \sqrt{1-\epsilon^2} f_b}Z'_\pm &=0, \label{eq:eomz} \\
    \ddot{Z}'_\pm + \left( k^2 + M_b^2 \pm \frac{k\dot{\phi}}{ f_b}\right)Z'_\pm + \frac{\epsilon \dot{\phi}}{ \sqrt{1-\epsilon^2} f_b}Z_\pm &=0. \label{eq:eomzp}
\end{align}
In order to find the tachyonic solutions of Eqs.(\ref{eq:eomz}) and (\ref{eq:eomzp}), we write the equations of motion in the matrix form as follow
\begin{align}
    \begin{pmatrix}
    \ddot{Z_\pm} \\ 
    \ddot{Z}'_\pm
    \end{pmatrix} = -
    \begin{pmatrix}
     k^2 + M_a^2 \pm \frac{\left(\epsilon^2 f_a + f_b \right)k\dot{\phi}}{ \left(1-\epsilon^2\right) f_a f_b} & \pm \frac{\epsilon k \dot{\phi}}{ \sqrt{1-\epsilon^2} f_b} \\
    \pm \frac{\epsilon k \dot{\phi}}{ \sqrt{1-\epsilon^2}f_b} & k^2 + M_b^2 \pm \frac{k\dot{\phi}}{ f_b}
    \end{pmatrix}
    \begin{pmatrix}
    Z_\pm \\
    Z'_\pm
    \end{pmatrix}. \label{eq:matrixz}
\end{align}
The above $2\times2$ matrix contains a negative eigenvalue when its determinant is less than zero. Therefore, the gauge boson will be exponentially produced when the following condition is satisfied:
\begin{equation}
    \frac{\epsilon^2k^2\dot{\phi}^2}{\left(1-\epsilon^2\right)f_b^2}\geq\left( k^2 + M_a^2 \pm \frac{\left(\epsilon^2 f_a + f_b \right)k\dot{\phi}}{ \left(1-\epsilon^2\right) f_a f_b} \right)\left( k^2 + M_b^2 \pm \frac{k\dot{\phi}}{ f_b}\right)>0. \label{eq:before_pp}
\end{equation}
In the large momentum limit, the particle production would be inhibited by the kinematics and hence the process happens inefficiently. Thus, we will consider only the low momentum limit, then the inequality becomes
\begin{equation}
    \dot{\phi}^2 > \frac{4\left(\epsilon^2-1\right)^2f_a^2f_b^2M_a^2M_b^2\left(M^2_a+M^2_b\right)}{f_b^2M^4_b+f_a^2\left( \left(\epsilon^2-1\right) M^2_a-\epsilon^2M^2_b \right)^2+2f_a f_b M^2_b\left( \left(\epsilon^2-1 \right)M^2_a+\epsilon^2M^2_b \right)}. 
    \label{eq:ppcondition}
\end{equation}
This condition is called the mixing particle production. We argue that the mixing particle production could happen at a significantly different scale comparing to the usual particle production mechanism. In order to see this, considering the decoupled limit ($\epsilon \to 0$), the condition for tachyonic solution becomes 

\begin{equation}
    \dot{\phi}^2>\frac{4 f_a^2 f_b^2  M_a^2 M_b^2 (M_a^2+M_b^2)}{\left(f_a M_a^2-f_b M_b^2\right){}^2}.\label{pdnomixing}
\end{equation}
Comparing with the thresholds for a set of 2 independent particle production conditions, we have
\begin{align}
    \dot{\phi}^2 > 4 f_a^2 M_a^2,\quad \text{and} \quad 
    \dot{\phi}^2 > 4 f_b^2 M_b^2. \label{eq:ordinarypp}
\end{align}
In the limit $M_a \sim M_b$ and $f_a \sim f_b$, the threshold for Eq.(\ref{pdnomixing}) could be much higher than Eq.(\ref{eq:ordinarypp}). On the other hand, in the limit $f_a M_a^2 \gg f_b M_b^2$, the mixing particle production threshold is comparable to the larger bound, i.e., $\frac{4 f_a^2 f_b^2  M_a^2 M_b^2 (M_a^2+M_b^2)}{\left(f_a M_a^2-f_b M_b^2\right){}^2} \sim 4f_b^2M_b^2$ 

\begin{figure}[t!]%
    \centering
    \includegraphics[height=7cm]{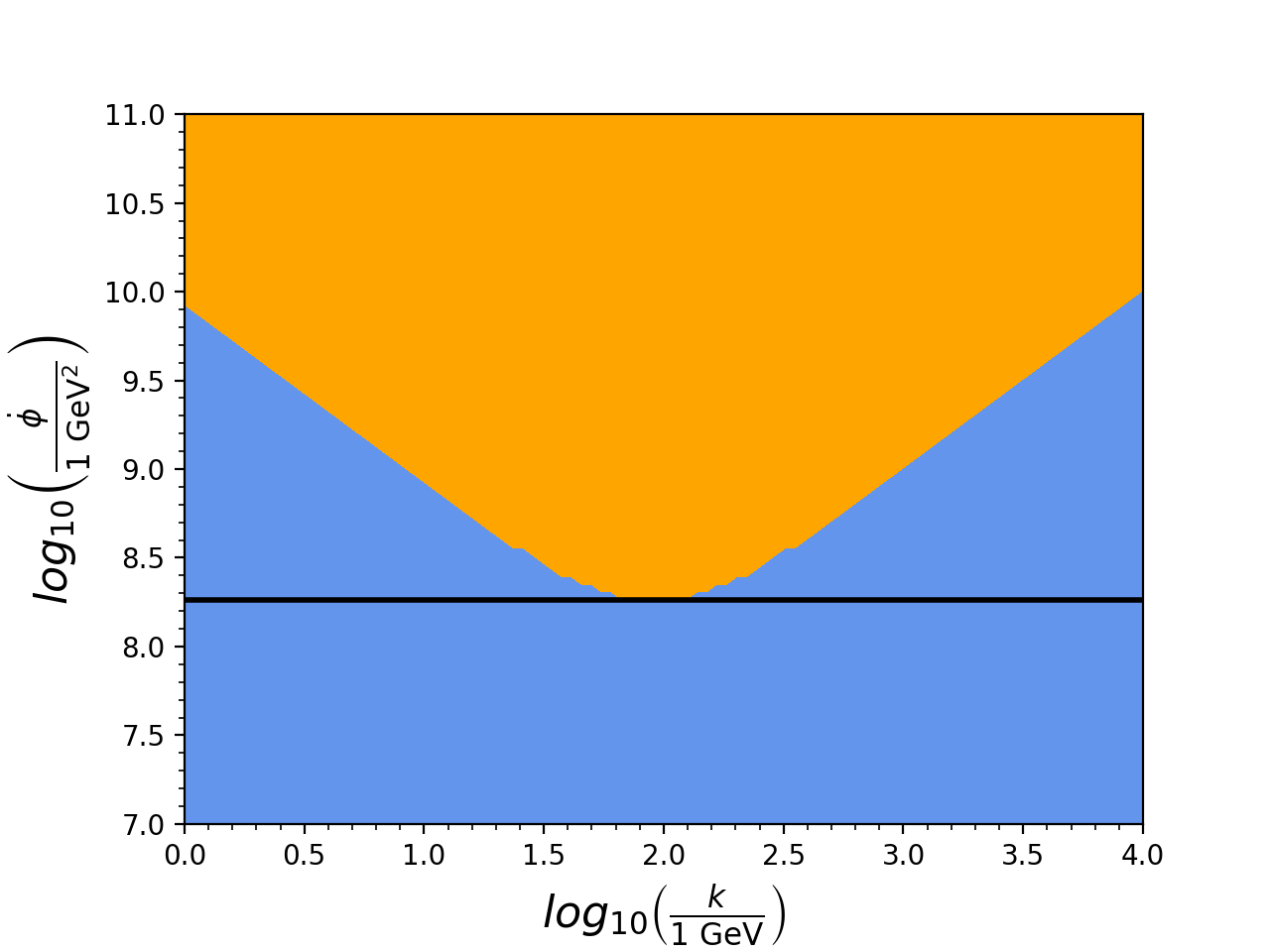} 
    \caption{The plot shows where the matrices in Eq.(\ref{eq:matrixz}) contain no negative eigenvalue (blue region) or at least one negative eigenvalue (orange region) in terms of the energy $\dot{\phi}$ and momentum $k$. This implies that the orange region is  where the mixing particle production condition is satisfied. The black line indicates the minimum energy $\dot{\phi}$ that the mixing particle production condition could be satisfied at corresponding momentum $k$, i.e., Eq.(\ref{eq:ppcondition}). Noted that the other parameters are used as $M_a=91\ {\rm GeV}, M_b=31622\ {\rm GeV}, f_a=10^6\ {\rm GeV}, f_b=10^6\ {\rm GeV}$ and $\epsilon=10^{-5}$.}
    \label{fig:ein}
\end{figure}

In order to verify the bound, we perform the full numerical analysis of the particle production condition and compare it with the mixing particle production condition. For each polarization, the eigenvalues of the matrix are calculated. The result is shown in Fig.(\ref{fig:ein}). The orange region indicates the value of momentum in which the matrices in Eq.(\ref{eq:matrixz}) contains at least one negative eigenvalue. Notice that the condition Eq.(\ref{eq:ppcondition}) is a good approximation since it has momentum modes in the orange region.

Due to the presence of light leptons coupled with gauge bosons, one could expect an interplay between the gauge bosons production and the Schwinger process of fermions \cite{Domcke:2021yuz}. As shown in \cite{Domcke:2018eki,Domcke:2019qmm}, the induced current is written as 
\begin{equation}
    J_f = \frac{\left(e |q_f| \right)^2}{6\pi^2}\frac{EB}{H}\coth{\frac{\pi B}{E}} \exp{\left(-\frac{\pi m^2_f}{e |q_f| E}\right)},
\end{equation}
where $q_f$ is an effective charge of the fermion $f$ under the gauge bosons and $E, B$ are absolute value of the gauge boson electric and magnetic fields. As we will see later in the scalar sector, at the scale at which the gauge boson production happens, the fermion mass is non-vanishing ($m_f \neq 0$) due to the Higgs vacuum expectation value.
Therefore, the suppression from Schwinger production of fermion is not dominant over our mixing production mechanism.
\section{Model Implementation} \label{sec:model}

In this section, we implement the mixing particle production to a dynamically breaking $U(1)'$ model. The additional field ingredients are a $U(1)'$ gauge boson ($Z'$ boson), a relaxion, and a new scalar responsible for the $Z'$ boson mass.


\subsection{The model}

In the scalar sector, we introduce the new scalar field, $\eta$, charged under $U(1)'$. The axion responsible for the relaxion mechanism, $\phi$, is also introduced. The Lagrangian for the scalar section is given by
\begin{eqnarray}
    \mathcal{L_{\rm scalar}}&=& D_\mu \eta D^\mu \eta + \left( D_\mu h\right)^\dagger \left( D^\mu h\right) + \left(\Lambda^2 - \kappa\phi\right)h h^\dagger + \left(\Lambda'^2 - \kappa'\phi\right)\eta\eta^\dagger + \kappa \Lambda^2\phi \nonumber\\
    && - \lambda h^4 - \lambda' \eta^4 + \lambda_\text{mix}h^2\eta^2 - V_{\rm np}(\phi), \label{eq:modellagrangian}
\end{eqnarray}
where $\kappa$ and $\kappa'$ are dimensionful parameters breaking the shift symmetry and $\Lambda$, $\Lambda'$ are the UV cutoff scale of the scalars. The $V_{\rm np}$ potential is generated from non-perturbative effects which is used as the barrier for the relaxion. The explicit breaking of the shift symmetry, i.e., the linear potential and the relaxion-scalar couplings, is utilised to enable the relaxation mechanism for $\eta$ and Higgs. In addition to the usual gauge sector Lagrangian, the anomaly couplings are written by
\begin{equation}
    \mathcal{L_{\rm gauge}} = -\frac{\phi}{4f}\left(g_2^2 W_{\mu\nu}^a \widetilde{W}^{a\;\mu\nu} - g_1^2 B_{\mu\nu}^a \widetilde{B}^{a\;\mu\nu}\right) - \frac{\phi}{4f'} Z'_{\mu\nu} \widetilde{Z}'^{\mu\nu}. \label{eq:gauge}
\end{equation}
The expansion of the first term in the mass eigenbasis gives
\begin{equation}
    -\frac{\phi}{f}\epsilon^{\mu\nu\rho\sigma}\left(2g_2^2 \partial_{\mu} W_{\nu}^- \partial_{\rho} W_{\sigma}^+ + (g_2^2 - g_1^2) \partial_{\mu} Z_{\nu} \partial_{\rho} Z_{\sigma} - 2 g_1 g_2 \partial_{\mu} Z_{\nu} \partial_{\rho}A_{\sigma}\right). \label{eq:gaugeexpand}
\end{equation}
The nonabelian part, i.e., the $WW$ term is suppressed where the $ZA$ term is also subdominant with respect to the $ZZ$ term. These terms also generate anomaly couplings to standard model fermions and photon as follows \cite{Bauer:2017ris,Craig:2018kne}
\begin{equation}
    \frac{\partial_{\mu}\phi}{f_F}\left(\overline{\psi}\gamma^{\mu}\gamma_5 \psi \right) + \frac{\phi}{4f_{\gamma}}F_{\mu\nu}\widetilde{F}^{\mu\nu},
\end{equation}
where
\begin{eqnarray}
    \frac{1}{f_F} &=& \frac{3\alpha_{\rm EM}^2}{4f}\left[ \frac{Y^2_{F_L} + Y^2_{F_R}}{\cos^4\theta_W} - \frac{3}{4\sin^4\theta_W}\right] \log\left(\frac{\Lambda_c^2}{m_W^2}\right),\\
    \frac{1}{f_{\gamma}} &=& \frac{2\alpha_{\rm EM}}{\pi \sin^2 \theta_W f} B_2(x_W) + \sum_{F}\frac{N_c^F Q_F^2}{2\pi^2 f_{F}}B_1(x_F),
\end{eqnarray}
where $Y_F$ is the hypercharge of the fermion field, $N_F$ is the color factor, $Q_F$ is the electric charge of the fermion. The funcions $B_1, B_2$ are defined as
\begin{eqnarray}
    &B_1(x) = 1 - xf(x)^2,\quad B_2(x) = 1 - (x-1)f(x)^2,&\\
    &f(x) = \begin{cases}
    \arcsin \left(\frac{1}{\sqrt{x}}\right) & x \geq 1\\
    \frac{\pi}{2} + \frac{i}{2}\log \left( \frac{1+\sqrt{1-x}}{1-\sqrt{1-x}}\right) & x <1
    \end{cases}&,
\end{eqnarray}
where $x_i = 4m_i^2/m_{\phi}^2$. In the limit of light axion $m_{\phi} \rightarrow 0$, $B_1(x_F) \sim m_{\phi}^2/\left(12m_F^2\right)$ and $B_2(x_W) \sim m_{\phi}^2/\left(6 m_W^2\right)$. Therefore, the exponential production of photon from the loop-induced couplings are negligible.

We propose that $\eta$ is charged under the $U(1)'$ gauge symmetry, therefore the $Z'$ obtains the mass from the spontaneous symmetry breaking of the $U(1)'$ symmetry and its mass is given by 
\begin{equation}
    M_{Z'} = M_b = \frac{g_X \langle \eta \rangle}{\sqrt{2}}, \label{eq:masszp}
\end{equation}
where $g_X$ is the gauge coupling of the $U(1)'$. On the other hand, the usual mass of $Z$ boson is given by the Higgs vev as
\begin{equation}
    M_{Z} = M_a = \frac{\sqrt{g_1^2 + g_2^2} \langle h \rangle}{\sqrt{2}}. \label{eq:massz}
\end{equation}
Since both $Z$ and $Z'$ boson masses are related to the dynamics of relaxion via the scalar vevs, we can use them in the triggering condition in Eq.(\ref{eq:ppcondition}). Comparing Eqs.(\ref{eq:gauge}) and (\ref{eq:gaugeexpand}) to the previous analysis we can identify
\begin{equation}
    f_a = \frac{f}{g_2^2 - g_1^2}, \quad f_b = f'.
\end{equation}
The lower bound of $\dot{\phi}^2$ for the mixing particle production becomes
\begin{equation}
    \frac{2 \left(\epsilon^2-1\right)^2 f^2 f'^2 g^2 \langle\eta\rangle ^2 g_X^2 \langle h\rangle^2 \left(g^2 \langle h\rangle^2 + g_X^2\langle\eta\rangle^2 \right)}{\left(\epsilon^2-1\right)^2 f^2 g^4 \langle h\rangle^4 - 2 \left(\epsilon^2-1\right) f g^2 \langle\eta\rangle^2 g_X^2 \langle h\rangle^2 \left(\epsilon^2 f - f' g'^2\right)+ g_X^4  \langle\eta\rangle^4 \left(\epsilon^2 f+ f' g'^2\right)^2},\label{eq:newcondition}
\end{equation}
where $g^2 = g_1^2 + g_2^2$ and $g'^2 = g_2^2 - g_1^2$.

From the Lagrangian Eq.(\ref{eq:modellagrangian}), the vev of $h$ and $\eta$ can be written as the following form:
\begin{align}
    \langle h \rangle =&\sqrt{\frac{2 \lambda' \left(\Lambda^2-\kappa  \phi_0 \right)+\lambda _\text{mix} \left( \Lambda'^2 - \kappa' \phi_0\right)}{4 \lambda  \lambda'-\lambda _\text{mix}^2}} , \label{eq:vevh} \\
    \langle \eta \rangle =&\sqrt{\frac{2 \lambda  \left(\Lambda'^2-\kappa' \phi_0\right)+\lambda _\text{mix} \left( \Lambda ^2 - \kappa  \phi_0  \right)}{4 \lambda  \lambda '-\lambda _\text{mix}^2}} ,
\end{align}
where $\phi_0$ is the value of the relaxion field at the given point in time, therefore the final value of $\phi_0$ determines the $\langle h \rangle$ and $\langle \eta \rangle$ after the relaxion stops. 
Since the $Z'$ mass must be higher than $Z$ mass, we are looking for the condition where $\langle h \rangle<\langle \eta \rangle$. If we choose small $\lambda_\text{mix}$, we then obtain the following inequality
\begin{equation}
    \Lambda ^2 \lambda '+\lambda  \phi_0  \kappa '<\lambda  \Lambda'^2+\kappa  \phi_0  \lambda '. \label{eq:conditionscalar1}
\end{equation}
The eigenvalues of the mass matrix are approximated as
\begin{align}
    m_{\phi}^2 =&  \frac{\partial^2 V_{\rm np}(\phi)}{\partial \phi^2}, \\
    m_{h}^2 =& 4 \left(\Lambda'^2 - \kappa' \phi \right) + \frac{2 \lambda_\text{mix}\left(\Lambda^2 - \kappa \phi\right)}{\lambda} + \mathcal{O}\left(\lambda_\text{mix}^2\right), \label{eq:massh} \\
    m_{\eta}^2 =& 4 \left(\Lambda^2 - \kappa \phi \right) + \frac{2 \lambda_\text{mix}\left(\Lambda'^2 - \kappa' \phi\right)}{\lambda'} + \mathcal{O}\left(\lambda_\text{mix}^2\right),\label{eq:masseta}
\end{align}
where we have assumed that the cut off scales are much higher than the shift symmetry breaking scales of the spurions, i.e., $\kappa \sim \kappa' \ll \Lambda \sim \Lambda'$ and $\lambda_{\rm mix} \ll 1$ is assumed in order to suppress the mixing angle. The full mass matrix and the derivation of its eigenvalues are given in appendix \ref{sec:massE}. Due to the small value of $\kappa$'s the mixing between Higgs or $\eta$ and the relaxion is negligible. On the contrary, the original model of relaxion utilising the feedback mechanism could lead to a significant mixing between Higgs and the relaxion, since the barrier becomes larger as the Higgs vev grows ($\Lambda \propto \langle h \rangle^j$) \cite{Flacke:2016szy}.
However, depending on the value of $\lambda_{\rm mix}$, we are left with only the mixing between Higgs and the new scalar, $\eta$. The mixing angle, $\theta$, is written as
\begin{align}
  \tan(2\theta) = \frac{\lambda_\text{mix}\sqrt{\left(\lambda_\text{mix} \left( \Lambda^2 - \kappa\phi \right) + 2 \lambda \left( \Lambda'^2 - \kappa'\phi \right) \right)\left(\lambda_\text{mix} \left( \Lambda'^2 - \kappa'\phi \right) + 2 \lambda' \left( \Lambda^2 - \kappa\phi \right) \right)}}{2\left( \lambda'\lambda_\text{mix}\left( \Lambda^2 - \kappa \phi \right) - \lambda\lambda_\text{mix}\left( \Lambda'^2 - \kappa' \phi \right) - 2\lambda\lambda'\left( \Lambda^2 - \Lambda'^2  + \left( \kappa' - \kappa \right)\phi \right) \right)}.
\end{align}
Since the Higgs couplings are in agreement with the standard model~\cite{Khachatryan:2016vau,CMS:2018uag,ATLAS:2019nkf}, we expect the scalar mixing angle to be small. In order to quantify a deviation from the standard model, the Higgs coupling strength modifier is defined as
\begin{equation}
	\mu_i^f = \frac{\sigma_i}{\sigma_i^{(SM)}}\frac{Br^f}{Br^f_{(SM)}},
\end{equation}
where $i$ indicates the production channel and $j$ indicates the decay channel. From~\cite{CMS:2018uag,ATLAS:2019nkf}, the global value is $\mu = 1.11 \pm 0.09$. Since all Higgs couplings are modified by $\cos{\theta}$, the predicted value is proportional to $\cos^4\theta$. Therefore, the consistency with the Higgs precision experiments requires
\begin{equation}
|\cos\theta| \geq \left(\frac{1}{1.11}\right)^{1/4} \geq 0.974.
\end{equation}

\subsection{Axion Relaxation}

\begin{figure}[h]%
    \centering
    \includegraphics[height=5.5cm]{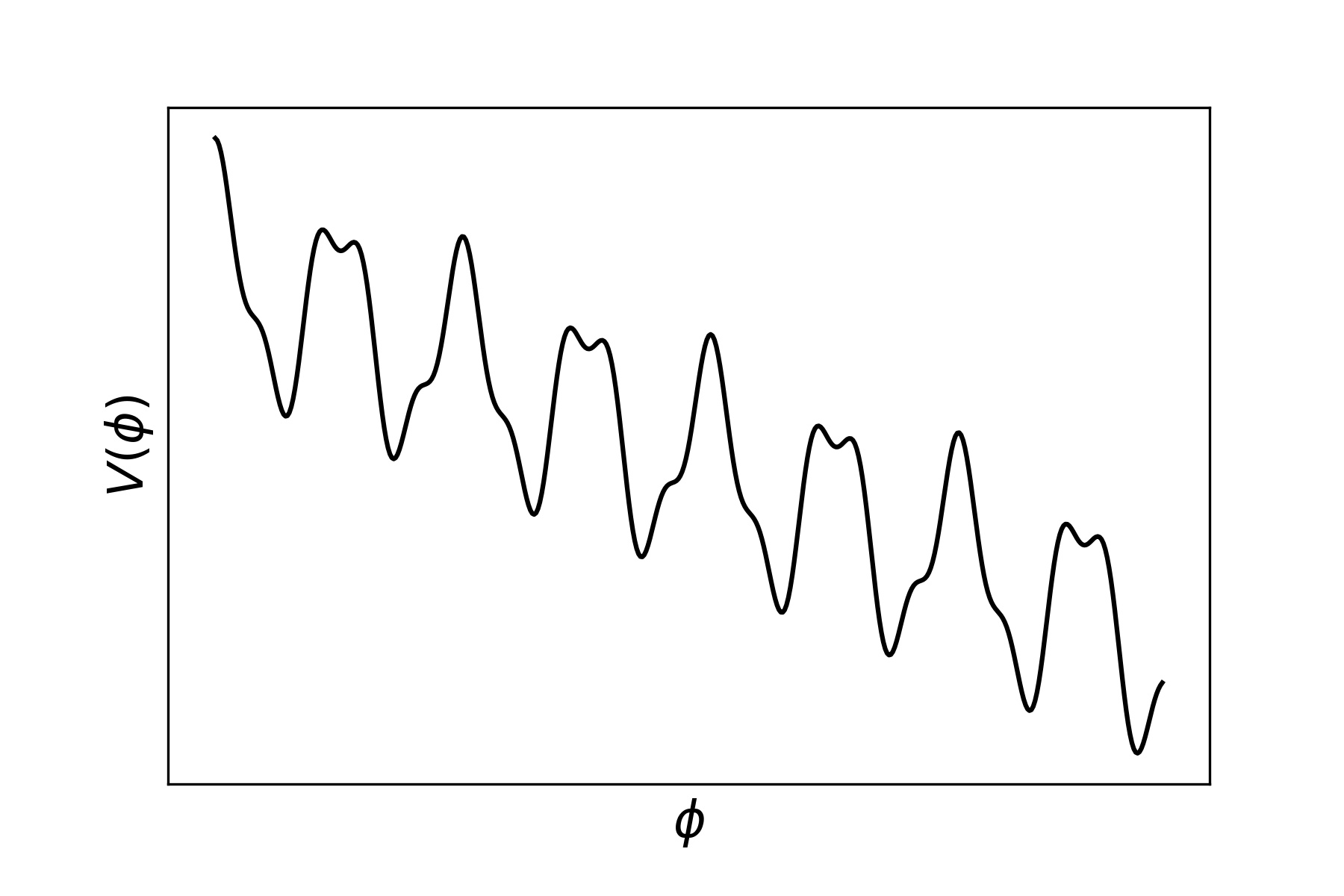}
    \caption{The sketch of the potential given in Eq.(\ref{eq:poten}). After the axion relaxation starts, relaxion field $\phi$ rolls down and produces the particles in particle production state until get trapped in one of the wiggles.}
    \label{fig:potentail}%
\end{figure}

Next let's discuss the relaxation mechanism in the model. As motivated by the UV completion perspective, we assume the simplest yet non-trivial form of potential, i.e., there are 2 sources of non-perturbative effects from some other gauge groups. Below their confinement scales, the axion-like potential can be approximated as 2 cosine terms. Combining with the linear part of the potential from Eq.(\ref{eq:modellagrangian}), we have 
\begin{equation}
    V = \Lambda_c^4 \cos\left(\frac{\phi}{F}\right) + {\Lambda'_c}^4 \cos\left(\frac{\phi}{F'}\right)-\kappa \Lambda^2\phi, \label{eq:poten}
\end{equation}
where $\Lambda_c$ and $\Lambda'_c$ are non-perturbative scales and $F$ and $F'$ are the axion couplings arising from the non-perturbative effects. The sketch of the potential can be seen in Fig.(\ref{fig:potentail}).

In order to have multiple minima in the potential, the slope of the linear part is required to be smaller than the slope of the barrier, i.e.,
\begin{equation}
    \kappa\Lambda^2 < \frac{\Lambda^4_c}{F} + \frac{\Lambda'^4_c}{F'}. \label{eq:condition1}
\end{equation}
Let's start with discussing the initial conditions, we first assume that the initial kinetic energy of $\phi$ is large enough to overcome the barriers, i.e.,
\begin{equation}
    \dot{\phi}^2_0 > \Lambda_c^4 + {\Lambda'_c}^4. \label{eq:condition2}
\end{equation}
In order to make it consistent to effective theory, we expect that $\Lambda^4>\dot{\phi}_0^2$. Therefore, we will assume that $\Lambda>\Lambda_c$. Since we expect that particle production is the mechanism to stop relaxion rolling, the relaxation has to occur after the inflation \cite{Fonseca:2019ypl}. In this scenario, the initial velocity of $\phi$ after inflation is required to be large enough in order to scan many vacua. 

After inflation ends ($H\sim \kappa$), the shift symmetry is broken and $\phi$ starts rolling. Since the masses of $h$ and $\eta$ start off with large and negative values, the initial field value should be around $\phi_0 \sim \Lambda^2/\kappa \sim \Lambda'^2/\kappa'$.
As shown in Eq.(\ref{eq:newcondition}), particle production occurs when
\begin{equation}
 \dot{\phi}^2 > \frac{2 \left(\epsilon^2-1\right)^2 f^2 f'^2 g^2 \langle\eta\rangle ^2 g_X^2 \langle h\rangle^2 \left(g^2 \langle h\rangle^2+\langle\eta\rangle^2 g_X^2\right)}{\left(\epsilon^2-1\right)^2 f^2 g^4 \langle h\rangle^4 - 2 \left(\epsilon^2-1\right) f g^2 \langle\eta\rangle^2 g_X^2 \langle h\rangle^2 \left(\epsilon^2 f - f' g'^2\right)+\langle\eta\rangle^4 g_X^4 \left(\epsilon^2 f+ f' g'^2\right)^2}. \nonumber
\end{equation}
At the beginning, the scalar vevs are large and the above inequality are not satisfied. When the effective scalar masses become sufficiently small, the gauge boson is exponentially produced which causes the dissipation of the kinetic energy of the relaxion. Eventually, the relaxion does not have enough energy to overcome the barrier and stop rolling at the vacuum that correspond to the correct Higgs mass.

Next, we discuss more requirements on the model. First, the vacuum energy must exceed the vacuum energy changes during the scanning of $\phi$, this requires
\begin{equation}
    H > \frac{\sqrt{\Lambda^4 + \Lambda'^4}}{m_{\text{pl}}}, \label{eq:condition3}
\end{equation}
where $H$ is Hubble scale and $m_\text{pl}$ is reduced Planck mass. The second is the requirement that the Hubble scale during inflation is lower than the cutoff scale so that the barriers form to stop the relaxion rolling,
\begin{equation}
    H < \text{Min}(\Lambda_c,\Lambda_c').
\end{equation}
Finally, the evolution of $\phi$ should be dominated by classical rolling. Therefore the rate of tunneling through the wiggles is smaller than the classical rate
\begin{equation}
    H < \left( \kappa \Lambda^2\right)^{1/3}.  \label{eq:condition6}
\end{equation}


\subsection{Axion Fragmentation} \label{ss:fragment}
\begin{figure}[h]%
    \centering
    \includegraphics[height=7cm]{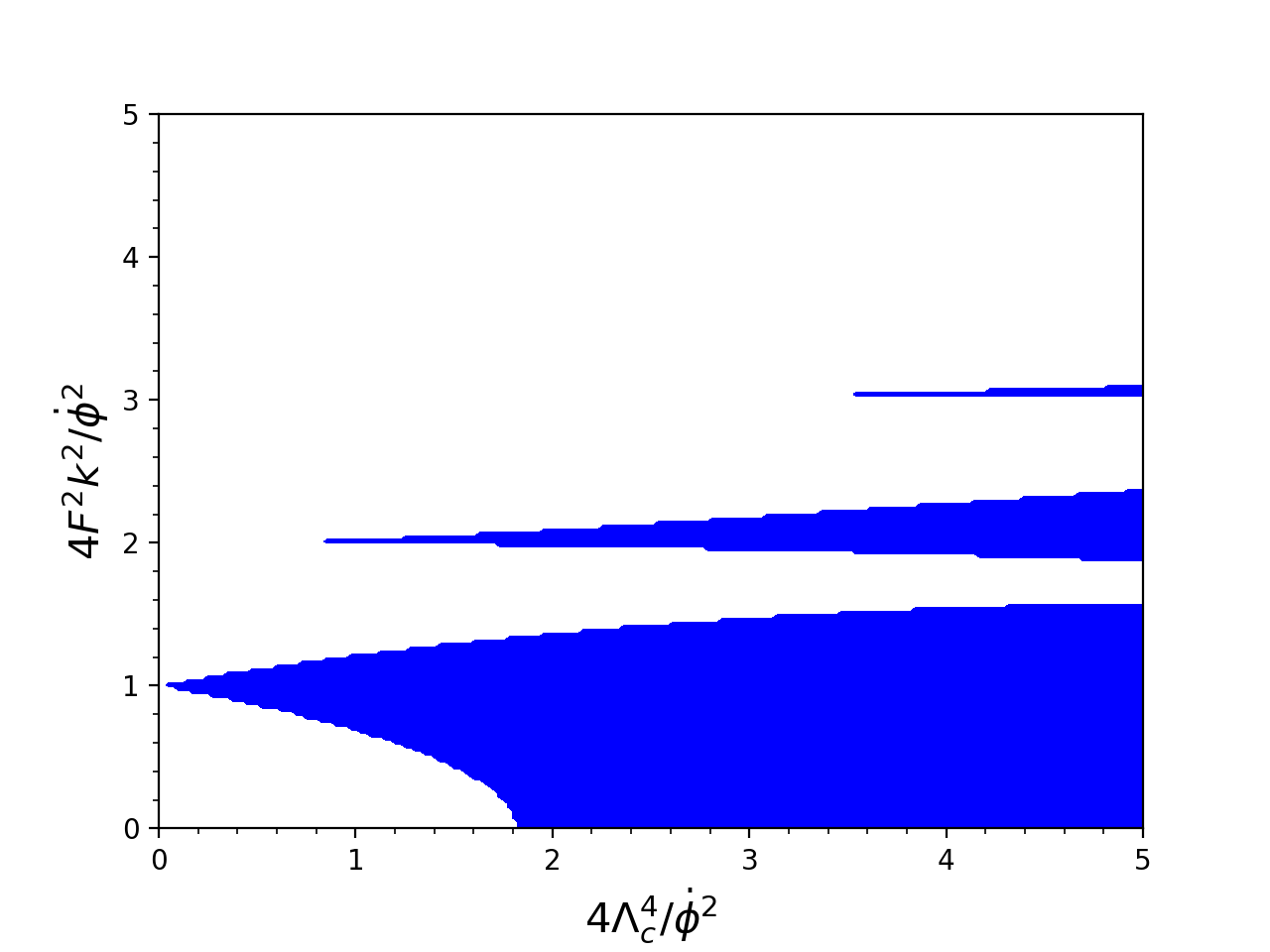}
    \caption{
    The plot shows the instability regions of Eq.(\ref{eq:QMH}) where 
    $\dot{\phi}=10^8\ {\rm GeV},F=10^9\ {\rm GeV},F'=10^{11}\ {\rm GeV}$, $\Lambda_c=\Lambda_c'=10^3\ {\rm GeV}$ and $\alpha=10^{-4}$ and $\nu=10^{-2}$.} Inside the blue area, the solution grows exponentially such that the axion fragmentation provides additional source of friction for relaxion stopping mechanism.
    \label{stab_region}%
\end{figure}

The effect of axion quantum fluctuations where the axion travels along a large number of minima has been investigated in \cite{Fonseca:2019ypl}. An additional source of friction provided by axion fragmentation is added and dominates the stopping mechanism of relaxion. It is worthwhile to explore this effect in our model. To do that, we decompose the relaxion field $\phi(x,t)$ into classical homogeneous mode, $\phi(t)$, and small fluctuation, $\delta\phi$, as
\begin{eqnarray}
    \phi(x,t)=\phi(t)+\delta\phi(x,t).
\end{eqnarray}
The equation of motion of $\phi(t)$ for the small fluctuation part can be written as mode function $u_k(t)$ as the following form,
\begin{equation}
    \frac{d^2u_k}{dt^2}+3H\frac{du_k}{dt}+\left[\frac{k^2}{a^2}+V''(\phi)\right]u_k=0,
    \label{eomdelphi}
\end{equation}
where $a$ is scale factor of the Friedmann–Lema{\^i}tre–Robertson–Walker metric and $H=\dot{a}/a$ is Hubble constant. By using the potential from Eq.(\ref{eq:poten}), in the limit where $\dot{\phi}=0$, $H = 0$ and $a=1$, the Eq.(\ref{eomdelphi}) is rewrite as
\begin{eqnarray}
    \frac{d^2u_k}{dt^2}+\left[k^2-\frac{\Lambda_c^4}{F^2}\cos\left(\frac{\dot{\phi}}{F}t\right)-\frac{\Lambda_c'^4}{F'^2}\cos\left(\frac{\dot{\phi}}{F'}t\right)\right]u_k=0.
    \label{eomdelphiDP}
\end{eqnarray}
The above equation is known as ``Quasiperiodic Mathieu-Hill Equation'' which generally present as following form,
\begin{eqnarray}
    u''+\left[\delta+\epsilon\left(\cos\left(2t
    \right)+\alpha\cos\left(2\nu t\right)\right)\right]u=0.
    \label{eq:QMH}
\end{eqnarray}
In our case, Eq.(\ref{eomdelphiDP}), each parameter in Eq.(\ref{eq:QMH}) can be defined as following
\begin{eqnarray}
    \delta&=&\left(\frac{2 \nu F' k}{\dot{\phi}}\right)^2=\frac{4F^2k^2}{\dot{\phi}^2},\quad \epsilon=-\frac{\Lambda^4_c}{F^2}\left(\frac{2 \nu F'}{\dot{\phi}}\right)^2=-\frac{4\Lambda^4_c}{\dot{\phi}^2},\nonumber \\
    \alpha&=&\frac{F^2}{F'^2}\frac{\Lambda_c'^4}{\Lambda_c^4}=\nu^2\frac{\Lambda_c'^4}{\Lambda_c^4},\quad
    \nu=\frac{F}{F'}. \label{eq:paraQP}
\end{eqnarray}
Although there were many analytical attempts in obtaining the stable region, only approximate results are found \cite{10.2307/2100807,10.1115/1.4037797,10.1115/1.4039144}. Hence in this work we present only the numerical study for the stable and unstable regions. The numerical solution for each point in the parameters space of $\left(\epsilon, \delta\right)$ is analysed with a random seed. In order to confirm its stability, the solution is checked against the time scale that a relaxion takes to travel from its initial point to the point where particle production happens, i.e., $\delta t \sim \phi_0/\dot{\phi}_0$. The region of stable and unstable solutions are shown in Fig.(\ref{stab_region}).

The exponential growth can be found in some regions of parameter space (also known as tongues). 
Note that the numerical results presented here contains only finite-time simulations which do not capture the infinite number of ``Arnold tongues'' permeating the stability region \cite{Braden:2010wd}. Since these infinite number of stability tongues are higher order effects leading to only a weak growing mode, the axion fragmentation effect in these Arnold tongues is expecting to be subdominant comparing to the main unstable regions. 
Numerical example of this weak resonance effect is given in Appendix \ref{app:Arnold}. 
However, due to the lowest order of the main instability regions, there is always some modes of an axion field containing instability which lead to the strong fragmentation effect.
It has been shown in \cite{Fonseca:2019ypl} that the fragmentation is effective ($\ddot{\phi} < 0$) when
\begin{equation}
    \kappa \Lambda^2 < 2H\dot{\phi}_0 + \frac{\pi \Lambda_c^8}{2F \dot{\phi}_0^2}\left(W_0\left(\frac{32\pi^2F^4}{e\dot{\phi}_0^2}\right)\right)^{-1},
\end{equation}
where $\dot{\phi}_0$ is initial velocity and $e$ is natural number or Euler's number. $W_0$ is the zeroth branch of the logarithm production function or Lambert $W$ function. 
\begin{equation}
    \kappa \Lambda^2 > \frac{\pi \Lambda_c^8}{F \dot{\phi}_0^2}\left(W_0\left(\frac{32\pi^2F^4}{e\dot{\phi}_0^2}\right)\right)^{-1}, \label{eq:conditionfrag}
\end{equation}
where we assume $\Lambda'_c = \Lambda_c$, $F = F'$. 

\section{Results} \label{sec:result}
From the previous sections, we have studied the constraints for the scalar sector, the relaxion mechanism, the particle production and the axion fragmentation. In this section, we provide examples of parameters that consistent with all the constraints listed above.
\begin{figure}[h!]%
    \centering
    \includegraphics[height=5.5cm]{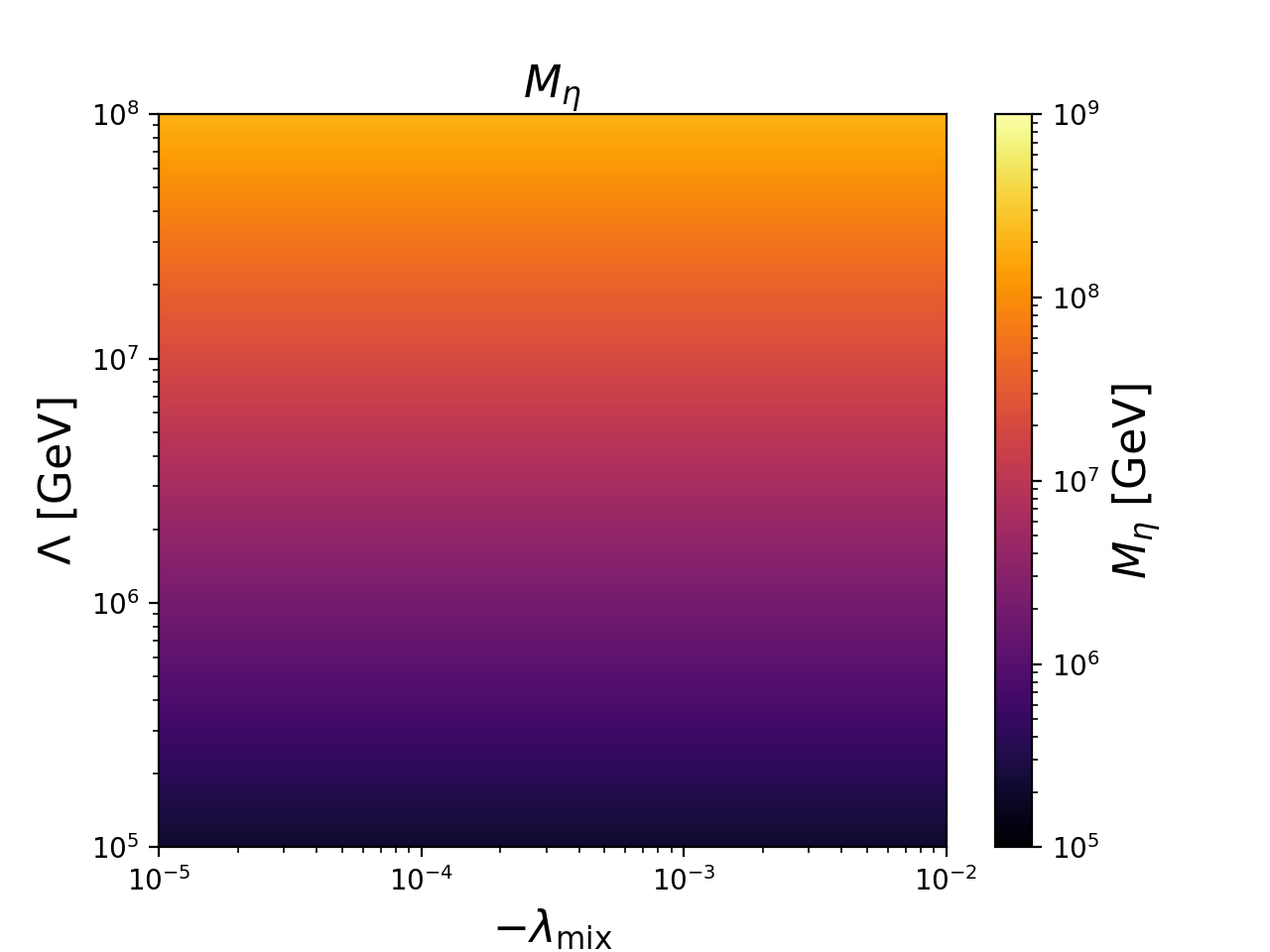} 
    \includegraphics[height=5.5cm]{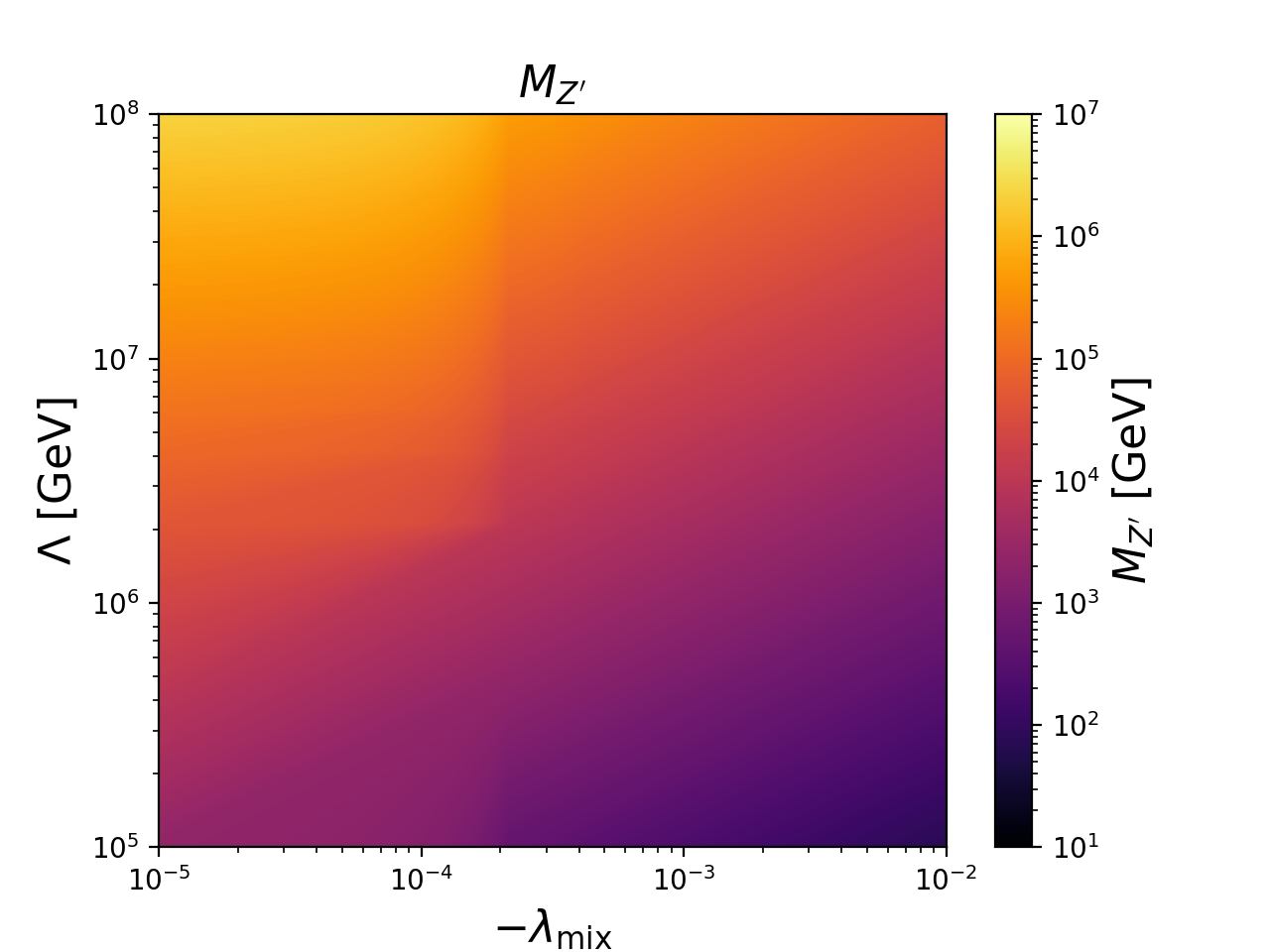} 
    \caption{
    The left contour represents the mass of $\eta$ field, $M_\eta$. The mass of $Z'$ boson, $M_{Z'}$, is shown on the right panel. Both contours are generated by using relevant parameters as following, $\Lambda=\Lambda'$, $\kappa=\kappa'=10^{-5}$, $\phi_0=10^3$, $\langle h \rangle=246 $ and $m_h=125$ GeV. Note that the coupling $g_X=10^{-4}$ is applied.} 
    \label{fig:contour1}
\end{figure}

For simplicity, we first assume that $\Lambda=\Lambda'$, $\kappa=\kappa'$, $\Lambda_c=\Lambda'_c$ and $f=f'$. Next, we fix the scalar parameters, $\lambda$ and $\lambda'$, by using Eqs.(\ref{eq:vevh}) and (\ref{eq:massh}) and the Higgs mass, $m_h=125$ GeV, and the Higgs vev, $\langle h \rangle=246 $ GeV. Note that, we only focus on the case that $\langle \eta \rangle > \langle h \rangle$ as this generally leads to $M_{Z'} > M_Z$. The example of $\eta$ mass and $Z'$ boson mass are shown in Fig.(\ref{fig:contour1}). It can be seen that the values of $M_{\eta}$ and $M_{Z'}$ can be made to be much higher than the typical current experimental constraints. In addition, we also found that the maximum mixing angle is generically less than the constraint $\cos{\theta}=0.974$.
In order to demonstrate an example set of parameters following the conditions (Eqs.(\ref{eq:condition1})-(\ref{eq:condition6}) and Eq.(\ref{eq:conditionfrag})), we choose $\lambda_\text{mix}=-10^{-3}$, $g=0.525$, $g_X = 10^{-3}$, $\epsilon = 10^{-5}$, and $\phi_0 = 10^3$ GeV.
The example of parameters is shown in Tab.(\ref{tab:tab1}). As a result, we found that the value of the parameters in our model are compatible with previous studies \cite{Fonseca:2018xzp}.

\begin{table}[h]
    \centering
    \begin{tabular}{|c|c|c|c|c|c|c|c|c|c|c|c|c|}
    \hline
    $\Lambda$ & $H$ & $\kappa$ & $\Lambda_c$ & $F$ & $F'$ & $f$ & $g'$ \\ \hline
    $10^4$ & $10^{-5}$ & $10^{-5}$ & $10^3$ & $10^9$ & $10^{10}$ & $10^6$ & $10^{-5}$ \\\hline
\end{tabular}
    \caption{The example set of parameters satisfying the condition of particle production mechanism after the inflation. Notice that we represent the parameters in GeV.}
    \label{tab:tab1}
\end{table}

\section{Conclusion and Discussion} \label{sec:conclusion}

In this paper, we consider the gauge bosons production as a friction mechanism to implement an axion relaxation for solving the electroweak hierarchy problem. We have shown that a linear combination of 2 heavy gauge bosons can be produced and the new mixing condition can be quantitatively different than the single heavy gauge boson case. We then present an implementation on the $U(1)'$ model. The new scalar field is introduced in order to separately break the gauge $U(1)'$ symmetry. The scalar sector containing Higgs, relaxion and the new scalar is analysed. Due to the Higgs vev independence of the barrier potential, we found the non-mixing nature of relaxion and Higgs particle. When the cufoff scale is sufficiently large, the mixing between Higgs and the scalar responsible for $U(1)'$ breaking is consistent with Higgs precision experiments. 

The relaxion mechanism and its constraints have been presented. The issue of the axion fragmentation where the small fluctuation of axion can experience a resonance effect, has been avoided by the high slope of the potential. Finally, an example for the viable set of parameters are presented. For the future work, we will explore the $U(1)'$ model and the mixing particle production mechanism from the UV perspective.





\section*{Acknowledgement}
The work of TK, CP and DS has been supported by the National Astronomical Research Institute of Thailand (NARIT). AW acknowledges the support of the Development and Promotion of Science and Technology Talents Project (DPST), the Institute for the Promotion of Teaching Science and Technology (IPST). DS is supported by the Mid-Career Research Grant 2021 from National Research Council of Thailand under contract no N41A640145. CP is supported by Research Grant for New Scholar, Office of the Permanent Secretary, Ministry of Higher Education, Science, Research and Innovation under contract no. RGNS 64-043. CP has also received funding support from the National Science, Research and Innovation Fund (NSRF).


\appendix
\section{Particle Production Condition}
In this section, we show detail calculations of the particle production in section \ref{sec:pp}. The matrix in Eq.(\ref{eq:matrixz}) can be written in a simpler form by defining new parameters as follows
\begin{align}
    A = \frac{\epsilon^2f_a + f_b}{(1-\epsilon^2)f_a f_b}, \quad 
    B = \frac{\epsilon}{\sqrt{1-\epsilon^2}f_b}, \quad
    C = \frac{1}{f_b}. \label{eq:newpara}
\end{align}
Therefore, the matrix in Eq.(\ref{eq:matrixz}) is rewritten as
\begin{align}
    \begin{pmatrix}
    \ddot{Z_\pm} \\ 
    \ddot{Z}'_\pm
    \end{pmatrix} = -
    \begin{pmatrix}
     k^2 + M_a^2 \pm A k\dot{\phi} & \pm B k \dot{\phi} \\
    \pm B k \dot{\phi} & k^2 + M_b^2 \pm C k\dot{\phi}
    \end{pmatrix}
    \begin{pmatrix}
    Z_\pm \\
    Z'_\pm
    \end{pmatrix}.
\end{align}
The particle production will occur when the solution of the above equation is an exponential form. This leads to
\begin{equation}
    \left(Bk\dot{\phi}\right)^2 \geq \left( k^2 + M_a^2 \pm A k\dot{\phi} \right)\left( k^2 + M_b^2 \pm C k\dot{\phi} \right). 
\end{equation}
In the low momentum limit $(k^3\approx0)$, the above condition can be simplified into the following form:
\begin{equation}
    k^2 \left( \left(AC-B^2\right) \dot{\phi} + M^2_a + M^2_b\right) \pm k\dot{\phi}\left( C M^2_a + A M^2_b\right) + M^2_a M^2_b \leq 0.
\end{equation}
One can rewrite it as
\begin{align}
    0\geq& \left(k\pm \frac{k\dot{\phi}\left( C M^2_a + A M^2_b\right)}{2 \left( \left(AC-B^2\right) \dot{\phi} +  M^2_a + M^2_b\right)} \right)^2 - \left( \frac{\dot{\phi}\left( C M^2_a + A M^2_b\right)}{2 \left( \left(AC-B^2\right) \dot{\phi} +  M^2_a + M^2_b\right)} \right)^2 \nonumber \\
    &+ \frac{M^2_a M^2_b}{\left( \left(AC-B^2\right) \dot{\phi} +  M^2_a + M^2_b\right)}.
\end{align}
Thus, the condition is satisfied when
\begin{equation}
    \left( \frac{\dot{\phi}\left( C M^2_a + A M^2_b\right)}{2 \left( \left(AC-B^2\right) \dot{\phi} +  M^2_a + M^2_b\right)} \right)^2
    - \frac{M^2_a M^2_b}{\left( \left(AC-B^2\right) \dot{\phi} +  M^2_a + M^2_b\right)} \geq 0,
\end{equation}
or
\begin{equation}
     \frac{\dot{\phi}^2\left( C M^2_a + A M^2_b\right)^2}{4 \left( \left(AC-B^2\right) \dot{\phi} +  M^2_a + M^2_b\right)}
    \geq M^2_a M^2_b. \label{eq:before_plug}
\end{equation}
Substitute Eq.(\ref{eq:newpara}) into Eq.(\ref{eq:before_plug}) and solve for $\dot{\phi}^2$, we obtain
\begin{equation}
        \dot{\phi}^2>\frac{4\left(\epsilon^2-1\right)^2f_a^2f_b^2M_a^2M_b^2\left(M^2_a+M^2_b\right)}{f_b^2M^4_b+f_a^2\left( \left(\epsilon^2-1\right) M^2_a-\epsilon^2M^2_b \right)^2+2f_a f_b M^2_b\left( \left(\epsilon^2-1 \right)M^2_a+\epsilon^2M^2_b \right)}.
\end{equation}

\section{Mass Eigenvalues \label{sec:massE}}
The mass matrix can be written as
\begin{align}
M = 
    \begin{pmatrix}
    M_{11} & M_{12} & M_{13} \\
    M_{21} & M_{22} & M_{23} \\
    M_{31} & M_{32} & M_{33} \\
    \end{pmatrix}, \label{eq:massmatrixfull}
\end{align}
where the components of the matrix are defined as
\begin{align}
    M_{11} =& 12 \lambda  \langle h\rangle ^2 - 2 \lambda_{\text{mix}} \langle \eta \rangle ^2 - 2 \Lambda ^2 + 2\kappa\phi , \\
    M_{22} =&  12 \lambda ' \langle \eta \rangle ^2 - 2 \lambda _{\text{mix}} \langle h\rangle ^2 - 2 \Lambda'^2 + 2\kappa'\phi , \\
    M_{33} =&  \frac{\Lambda _c^4}{F^2} \cos \left(\frac{\phi }{F}\right) + \frac{\Lambda'^4_c}{F'^2} \cos \left(\frac{\phi }{F'}\right) , \\
    M_{12} =& M_{21} = -2 \lambda _{\text{mix}} \langle \eta \rangle  \langle h\rangle , \\
    M_{13} =& M_{31} = \kappa  \langle h\rangle, \\
    M_{23} =& M_{32} = \kappa' \langle \eta \rangle .
\end{align}

Since $\kappa$ and $\kappa'$ are very small, we can assume that $\kappa$ and $\kappa'$ is zero. However, we cannot neglect them in diagonal terms since $\phi$ is very large comparing to $\kappa$ and $\kappa'$. The mass mixing matrix is then simplified as
\begin{align}
M = 
    \begin{pmatrix}
    M_{11} & M_{12} & 0 \\
    M_{21} & M_{22} & 0 \\
    0 & 0 & M_{33} \\
    \end{pmatrix}. \label{eq:massmatrix}
\end{align}
The eigenvalues of Eq.(\ref{eq:massmatrix}) are
\begin{align}
    m_{\phi}^2 =&  \frac{\Lambda _c^4 }{F^2}\cos \left(\frac{\phi }{F}\right) + \frac{\Lambda_c'^4 }{F'^2}\cos \left(\frac{\phi }{F'}\right), \\
    m_{h}^2  =& \frac{4\left( \lambda\lambda_\text{mix}\left( \Lambda'^2 - \kappa'\phi\right) + \lambda'\lambda_\text{mix}\left( \Lambda^2 - \kappa\phi\right) +2\lambda\lambda'\left( \Lambda^2 + \Lambda'^2 -(\kappa+\kappa')\phi\right) \right) +2\sqrt{D}}{4\lambda\lambda'-\lambda_\text{mix}^2}, \\
    m_{\eta}^2  =& \frac{4\left( \lambda\lambda_\text{mix}\left( \Lambda'^2 - \kappa'\phi\right) + \lambda'\lambda_\text{mix}\left( \Lambda^2 - \kappa\phi\right) +2\lambda\lambda'\left( \Lambda^2 + \Lambda'^2 -(\kappa+\kappa')\phi\right) \right)-2\sqrt{D}}{4\lambda\lambda'-\lambda_\text{mix}^2} ,
\end{align}
where
\begin{align}
    D =& \left( \lambda_\text{mix}^2-16\lambda\lambda'\right)\left( \lambda_\text{mix}\left(\Lambda^2 -\kappa\phi\right) + 2\lambda\left(\Lambda'^2 - \kappa'\phi\right) \right) \left( 2\lambda'\left(\Lambda^2 -\kappa\phi\right) + \lambda_\text{mix}\left(\Lambda'^2 - \kappa'\phi\right) \right)\nonumber \\
    &+4\left( \lambda'\lambda_\text{mix}\left(\Lambda^2 - \kappa\phi\right) + \lambda\lambda_\text{mix}\left(\Lambda'^2 -\kappa'\phi\right) + 2\lambda\lambda'\left(\Lambda^2 + \Lambda'^2 -(\kappa+\kappa')\phi\right) \right)^2 .
\end{align}

\section{Numerical Solutions of Arnold Tongues} \label{app:Arnold}
\begin{figure}[h]%
    \centering
    \includegraphics[height=5.4cm]{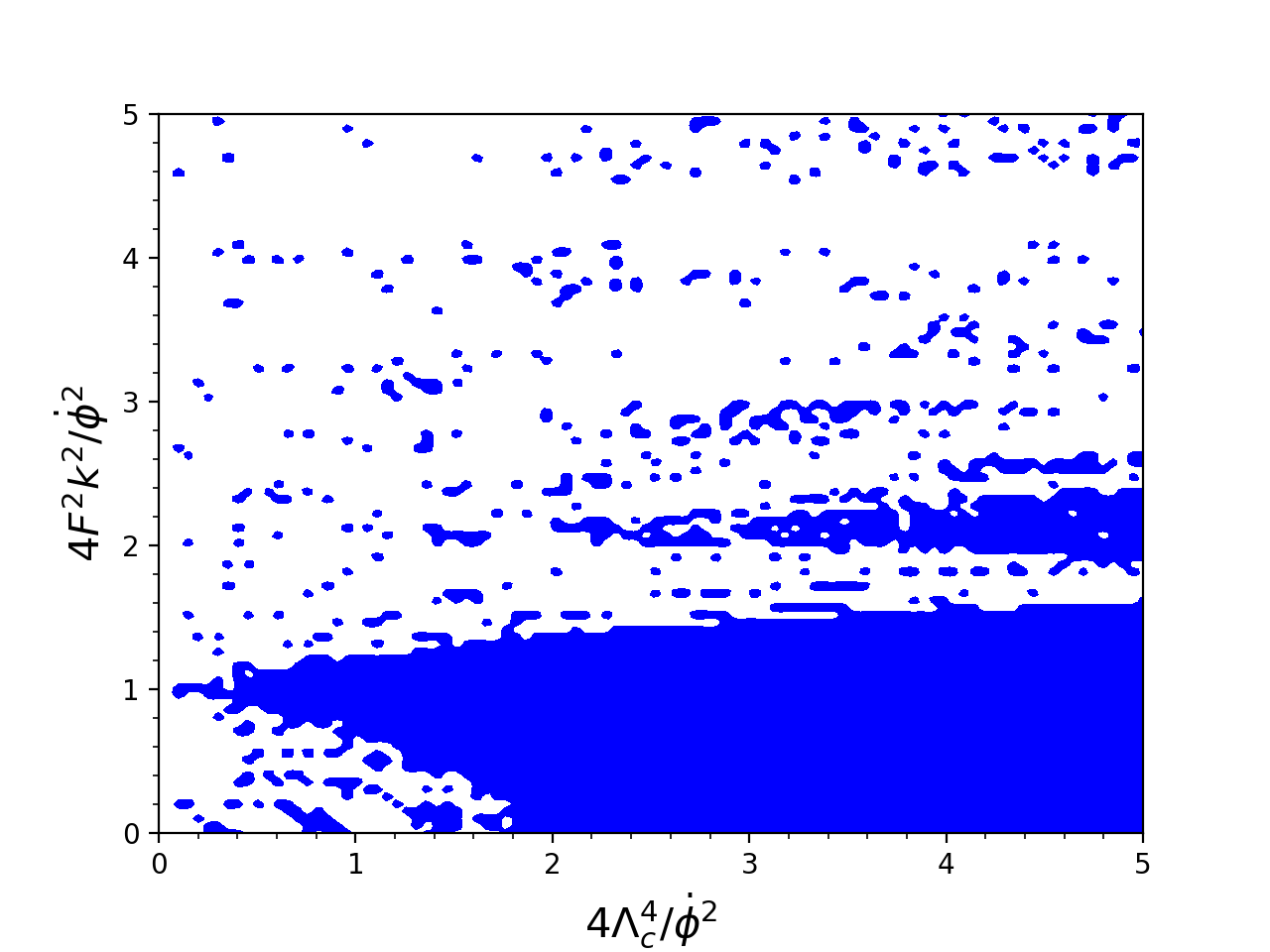}
    \includegraphics[height=5.2cm]{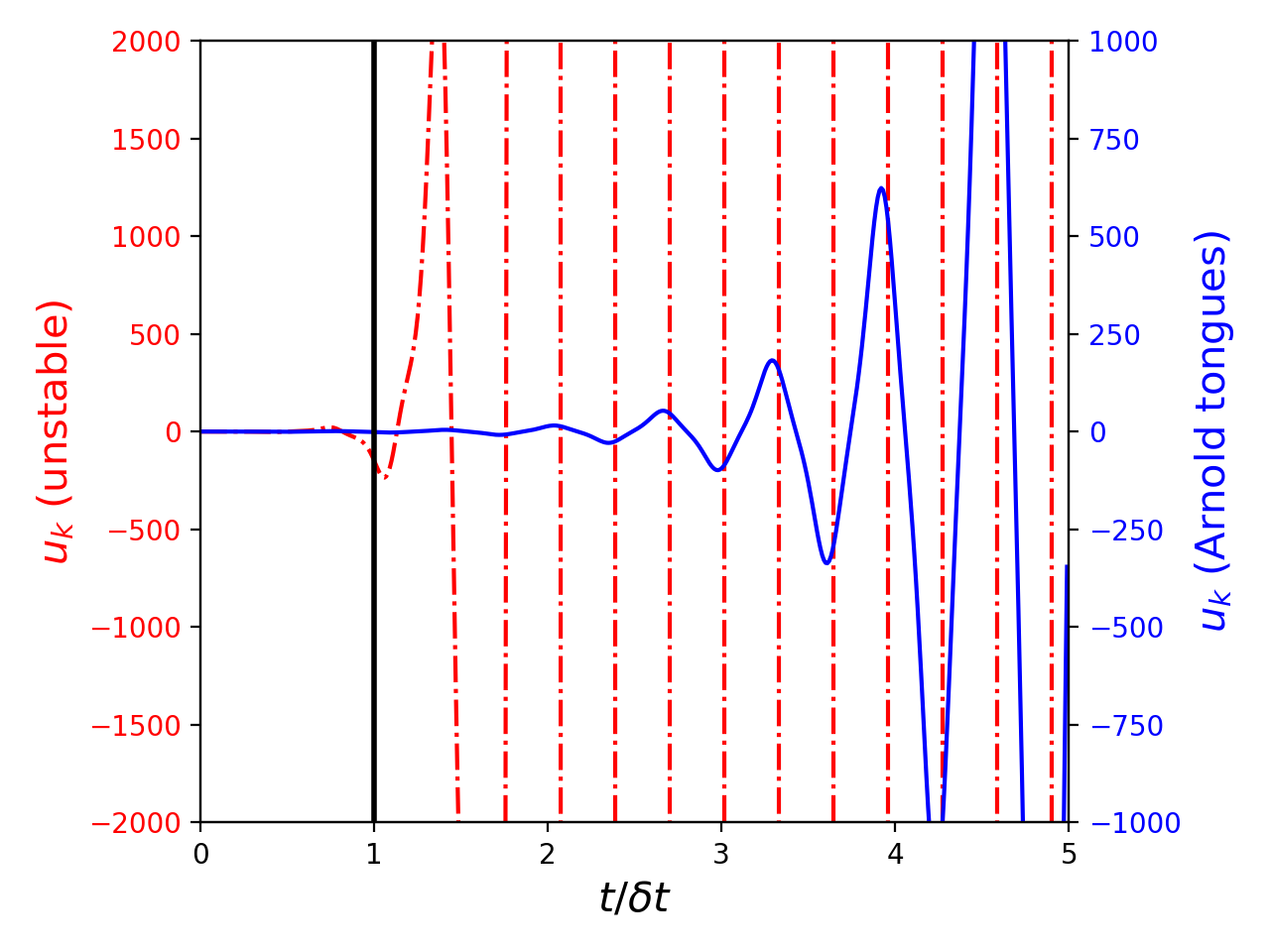}
    \caption{
    The left plot shows the instability band with Arnold tongues. The right plot shows the solution where the parameters come from the unstable band (red) and from the Arnold tongues region (blue). The black line is the time scale where the particle production starts, $\delta t \sim \phi_0/\dot{\phi}\sim 10$ GeV$^{-1}$. The parameters from the unstable band are chosen from the main tongue with $\frac{4 F^2 k^2}{\dot{\phi}^2} = 1.0 $ and $\frac{4\Lambda_c^4}{\dot{\phi}^2} = 4.0$. The Arnold tongue solution is given by $\frac{4 F^2 k^2}{\dot{\phi}^2} = 0.152$ and $\frac{4\Lambda_c^4}{\dot{\phi}^2} = 1.87$.}
    \label{unstable}
\end{figure}

In this section, we present the numerical analysis of the Arnold tongues discussed in subsection \ref{ss:fragment}. The presence of Arnold tongues is depicted in Fig.(\ref{unstable}) where the time scale of checking the stability is set to be $t = 15\delta t$, where $\delta t \sim \phi_0/\dot{\phi}_0 \sim 10^{9}/10^8 \;{\rm GeV}^{-1} \sim 10 \;{\rm GeV}^{-1}$ is the time axion takes to travel before the gauge bosons production. We also show the numerical solution for one of the points on Arnold tongues on the right hand side of Fig.(\ref{unstable}) comparing it with an unstable solution from the main tongue.
The solution shows the weak resonance effect in the Arnold tongues since the convergence time is larger than the time scale from the relaxion mechanism.




\bibliographystyle{jhep}
\bibliography{ref}

\end{document}